\documentstyle[prd,aps,preprint]{revtex}
\begin{document}
\draft

%
%

\preprint{Nisho-97/5} \title{Ferromagnetism of Axion Domain Wall} 
\author{Aiichi Iwazaki}
\address{Department of Physics, Nishogakusha University, Shonan Ohi Chiba
  277,\ Japan.} \date{January 29, 1997} \maketitle
\begin{abstract}
We show that axion domain wall is ferromagnetic in the universe with 
nonvanishing baryon number or lepton number. It is caused by 
protons and electrons bounded to the domain wall with their spins
aligned. 
These bound states arise due to attractive potentials generated 
through pseudo-vector couplings between the fermions and the axion.
Using a model of hadronic axions we predict 
existence of a primordial magnetic field 
with strength $10^{-13}$ Gauss at recombination.
\end{abstract}
\pacs{14.80.Mz, 98.80.Cq, 11.30.Er  
\hspace*{3cm}}
\vskip2pc

The axion\cite{Wilczeck} is a Nambu-Goldstone mode 
associated with Peccei-Quinn symmetry\cite{Pecci}
which leads to a possible solution of strong CP problem\cite{t'Hooft}. 
The axion gains
a mass through QCD instanton interactions. These interactions give rise 
to several possible vacuum expectation values of the axion field. 
This fact in general
leads to axion domain walls\cite{Sikivie,Turner} in cosmology. 
Namely in the early
universe the axion field can take different vacuum expectation values 
in different 
causally distinct regions when the temperature of the universe 
cools down to below QCD transition temperature. 
Thus the domain walls
appear between these regions. However the energy of the domain walls 
dominates the energy of the universe so quickly that 
the standard scenario of the early universe is invalidated\cite{Zeld,Turner}.
In order to solve the problem it is helpful to find new properties 
of the domain wall.

In this paper we reveal a new feature of the axion domain wall: The domain 
wall is ferromagnetic. This property is caused by polarized 
neucleons and leptons bounded to the wall. We will show that the domain wall
produces a potential in which only the particles polarized perpendicular
to the wall are bounded. The potential has the width of $m_a^{-1}$ and 
the depth of $m_a$ approximately where $m_a$ is the axion mass. 
Thus the binding energy is the order 
of the axion mass ( $10^{-3}$ eV$\sim10^{-5}$ eV ). As we consider
a period when the temperature of the universe is higher than that of 
neucleosynthesis, but lower than the temperature ( $150$ MeV$\sim 100$ MeV ) 
of QCD transition, relevant particles to the property of the wall are
protons, electrons, and their anti-particles. These particles form bound 
states in the potential where the spins of the states are aligned 
in the identical direction 
perpendicular to the wall. Thus as far as the baryon number 
( the lepton number ) of 
the universe is non-zero, the wall possesses magnetization just as 
ferromagnetics of ordinary matter does. 
This is a mechanism of ferromagnetism of the axion domain wall, which is 
different from a mechanism we have recently discussed \cite{iwazaki} in 
domain walls with fermion zero modes.

Now we show that the attractive potential for neucleons or leptons is 
produced by the axion domain wall. For the purpose we note that 
axion interaction\cite{kim} 
with these particles is specified as follow,

\begin{equation}
L_{int}=g\partial_{\mu}a \bar{\psi}\gamma_5\gamma^{\mu}\psi+h.c.
\end{equation}  
where $a$ is the axion field and $\psi$ is the fermion field ( for our 
purpose we suppose that the fermions are protons or electrons ). $g$ is 
the coupling constant whose value depends on models\cite{Kim,Dine} 
of the axion. We assume that for both hadrons and leptons it is 
the order of $f_{PQ}^{-1}$, the breaking scale of Peccei-Quinn symmetry ( 
$f_{PQ}=10^{10}$ GeV $\sim 10^{12}$ GeV )\cite{Turner}. When we consider 
a model of hadronic axion, the coupling constant with leptons is the order of 
$\alpha f_{PQ}^{-1}$; $\alpha=1/137$.

Suppose that the domain wall is located at $x_3=0$ and is extending in 
$x_1$ and $x_2$ directions. Then the nonvanishing component of 
$\partial_{\mu}a$ is $\partial_3 a$. 
Dirac equation $(i\gamma_{\mu}\partial^{\mu}-m + ga\gamma_3\gamma_5)\psi=0$
is reduced to

\begin{eqnarray}
((E-m)\sigma_3-g\partial_3a)u+i\partial_3v=0, \quad 
(-(E+m)\sigma_3+g\partial_3a)v-i\partial_3u=0
\end{eqnarray}
where $E$ $(m)$ is the energy ( mass ) of the fermion and we have set
$\psi= {u\choose v}$, assuming fermions with no momenta 
in $x_1$ and $x_2$ directions.
We have used a convention in the Bjorken-Drell text for gamma metrices (
$\sigma_3$ is Pauli metrics ). Taking account of smallness of $g\partial a$ 
in comparison with the fermion mass,
we obtain the Schr\"odinger equation for the large component $u$
( $v$ ) in the nonrelativistic limit,
 
\begin{equation}
\varepsilon u = (-\frac{\partial_3^2}{2m} + g\partial_3a\sigma_3)u
\quad 
\left(\ \varepsilon v = (-\frac{\partial_3^2}{2m} - 
g\partial_3a\sigma_3)v\ \right)
\label{S}
\end{equation}
with $E=m+\varepsilon$ ( $E=-m-\varepsilon$ ) 
where small component $v=-i\partial_3u/2m$ ( $u=i\partial_3v/2m$ ).

Note that $a$ is given by $f_{PQ}\theta$ ( $\theta$ is angle variable 
changing from $0$ to $2\pi$ around the domain wall ) and 
that the typical scale of spatial variation of $\theta$ is given 
by the axion mass $m_a$. Thus it turns out that 
the potential $g\partial_3a\sigma_3$ 
in eq(\ref{S}) for both of 
the fermion and the antifermion
is attractive for the state with spin down,
$u={0\choose 1}$ and repulsive for the state with spin up,
$u={1\choose0}$. It has approximately 
the width of $m_a^{-1}$ and the depth of $m_a$. $m_a$ is about
$10^{-3}$eV $\sim 10^{-5}$eV. Hence the potential depth is quite shallow but
the width is so large ( the order of $1$ cm ) that the potential can 
accommodate bound states for protons and electrons. 
In such the wide potential the particles move almost freely in $x_3$ direction.
Obviously, spins of these particles bounded to the wall are aligned.
This causes the ferromagnetism of the axion domain wall.

We may take the energies of the bound states approximately such that 
$\varepsilon=-m_a+k_3^2/2m$ where $k_3$ is the momentum in $x_3$ direction.
These bound states move also in $x_1$ and $x_2$ directions. 
Then the energies
of the bound states are specified with their momenta $\vec{k}$, 

\begin{equation}
\varepsilon= -m_a + \frac{\vec{k}^2}{2m}
\end{equation} 

In order to calculate the magnetization we need energy spectra of these states
in external magnetic field, $B$ perpendicular to the wall. It is easy to 
derive the energies of the fermions and antifermions in the nonrelativistic
limit,

\begin{eqnarray}
\varepsilon_{k,n+1}&=&-m_a + \frac{k_3^2}{2m} + 
\omega(n+1)\quad  \mbox{for fermion}\nonumber\\
\varepsilon_{k,n}&=&-m_a + \frac{k_3^2}{2m} +\omega n\quad 
 \mbox{for antifermion}
\end{eqnarray}
where $n$ is integer ( $\geq 0$ ) and 
$\omega$ is the cyclotron frequency
( $\omega=eB/m$ ).
The difference in the energy spectra 
between the fermions and the antifermions comes from the difference in 
the directions of the magnetic moments of these particles bounded to the wall.

Now we calculate a free energy of these particles bounded to the wall, 

\begin{equation}
\Omega=-\beta^{-1}\sum_{\varepsilon_{k,n+1}\leq 0}
\log{(1+ e^{(\mu-m-\varepsilon_{k,n+1})\beta})}-
\beta^{-1}\sum_{\varepsilon_{k,n}\leq 0 }
\log{(1+e^{-(\mu+m+\varepsilon_{k,n})\beta})}
\end{equation}
where the first term represents 
contributions of fermions, while 
second one does those of anti-fermions. 
$\mu$ is the chemical potential of the baryon number or the lepton number 
of the universe.
$\sum_{\varepsilon_{k,n+1}\leq 0}$ ( $\sum_{\varepsilon_{k,n}\leq 0 }$ ) 
implies $N_d\sum_{n=0}\int m_a^{-1}dk_3/(2\pi)$ 
with the condition 
that $\varepsilon_{k,n+1}\leq 0$ ( $\varepsilon_{k,n}\leq 0$ ); 
we sum up only contributions from the bound states. ( $N_d$ is the degeneracy
given by $eBL^2/2\pi$ where $L^2$ is the surface area of the wall ).

The magnetization, $M$ per unit volume of the wall is then given by, 

\begin{equation}
M=-\frac{1}{lL^2}\frac{\partial\Omega}{\partial B}
\end{equation}  
where $l$ is the width of the wall ( $l=m_a^{-1}$ ).

Let us first calculate the free energy 
contributed by protons and the corresponding magnetization 
in the case of the temperature 
being much lower than the proton mass. Because
we consider the case that the energy $\varepsilon$ is much lower 
than the temperature, we expand $\Omega$ in terms of $e^{-m_p\beta}$ and 
$\varepsilon\beta$ and only take the leading term,

\begin{equation}
\Omega=-\beta^{-1}e^{-m_p\beta}(\cosh\mu\beta 
 (\sum_{\varepsilon_{k,n+1}\leq 0}+\sum_{\varepsilon_{k,n}\leq 0})
+\sinh\mu\beta 
 (\sum_{\varepsilon_{k,n+1}\leq 0}-\sum_{\varepsilon_{k,n}\leq 0}))
\end{equation} 
where $m_p$ is proton mass.
The summation is performed as follows,

\begin{eqnarray}
\sum_{\varepsilon_{k,n+1}\leq 0}&=&N_d\sum_{n=0}\int \frac{ldk_3}{2\pi}=
\frac{N_dl\sqrt{2m_p\omega}}{\pi}(\sum_{n=0}^{n=F}
\sqrt{\frac{m_a}{\omega}-n}
-\sqrt{\frac{m_a}{\omega}}) \nonumber\\
&=&\frac{N_dl\sqrt{2m_p\omega}}{\pi}(\frac{2F^{3/2}}{3}+\frac{1}{2}F^{1/2}+
O(F^0)-
\sqrt{\frac{m_a}{\omega}}) \\
\sum_{\varepsilon_{k,n}\leq 0}&=&\frac{N_dl\sqrt{2m_p\omega}}{\pi}(
\frac{2F^{3/2}}{3}+\frac{1}{2}F^{1/2}+O(F^0))
\end{eqnarray}
where we have set $F=m_a/\omega$ and taken the limit of 
the small magnetic field, $F\to \infty$. Note that a linear term in $B$ arises 
only from the combination, 
$\sum_{\varepsilon_{k,n+1}\leq 0}-\sum_{\varepsilon_{k,n}\leq 0}$;
this term determines the magnetization.

Thus $M$ is given such that  

\begin{equation}
M=\beta^{-1}\sinh\mu\beta e^{-m_p\beta}\frac{e(2m_pm_a)^{1/2}}{2\pi^2}
\end{equation}
in the limit of the small magnetic field.
Here we rewrite the chemical potential $\mu$ in terms of the ratio, $N_B$ of 
the baryon number density to the entropy density of the universe\cite{Turner}; 
$N_B=n_B/s$ where
$n_B=8e^{-m_p\beta}\sinh{\mu\beta}(m_p/2\pi\beta)^{3/2}$ 
is the baryon number density and $s=2\pi^2g_*/45\beta^3$ is 
the entropy density. 
Then it follows that

\begin{equation}
M_p=\frac{e\pi^{3/2}g_*}{90}\frac{\sqrt{m_a}N_B}{\beta^{5/2}m_p}
\end{equation}
where $g_*$ is the number of massless degrees of freedom 
at the temperature $\beta^{-1}$ ( = $1\sim 100$ MeV ).

Similarly we calculate the contribution from electrons to
the magnetization of the domain wall. In the case the mass, $m_e$ of electron
is much lower than the temperature under consideration. 
Thus we expand the free energy in terms 
of $\mu\beta$, $m_e\beta$ and $\varepsilon\beta$,

\begin{eqnarray}
\Omega&=&-\beta^{-1}(\log2-\frac{m_e\beta}{2}) 
( \sum_{\varepsilon_{k,n+1}\leq 0} + 
\sum_{\varepsilon_{k,n}\leq 0}) -\frac{\mu}{2}(\sum_{\varepsilon_{k,n+1}\leq 0}
-\sum_{\varepsilon_{k,n}\leq 0})\nonumber\\
&+&\frac{(\sum_{\varepsilon_{k,n+1}\leq 0} + 
\sum_{\varepsilon_{k,n}\leq 0})\varepsilon_{k,n+1}}{2}
\end{eqnarray} 
where we have taken the leading term of the order of $\beta^{-1}$ and 
the next leading term of the order of $\mu$ and $\varepsilon$. 
We have also taken the limit of the small magnetic field;
$m_e$ is electron mass. 
Then, it follows that $M_e=e\mu\sqrt{2m_em_a}/4\pi^2$. 
Rewriting the chemical potential in terms of 
the ratio $N_L=15\mu\beta/2\pi^2g_*$ of the lepton 
number density to the entropy density, we obtain

\begin{equation}
M_e=\frac{e\sqrt{2m_em_a}g_*N_L}{30\beta}
\end{equation}

We find that both of the magnetization $M_p$ and $M_e$ are proportional to
the baryon number $N_B$ and the lepton number $N_e$, respectively.
This is resulted from the fact that the magnetization arises owing to 
polarized particles and antiparticles bounded to the wall;
their spins are aligned in the identical direction so that 
the sum of their magnetic moments are proportional to $N_B$ or $N_L$.

We see that the magnetization $M_p$ decreases much faster 
with the temperature than the magnetization $M_e$ does. 
Thus as far as we consider the temperature below 
$100$MeV, the magnetization of the axion domain wall is mainly determined by 
the effect of electrons if $N_B=N_L$. 
Numerically we find that 

\begin{equation}
M\sim 10^{-1}\sqrt{\frac{m_a}{10^{-4}\mbox{eV}}}
\frac{\beta^{-1}}{100\mbox{MeV}}\frac{N_L}{10^{-10}}\mbox{Gauss}
\end{equation}
with $g_*\sim 10$.
This formula holds in the temperature below $100$MeV and above $1$MeV.
( When we consider the hadronic axion, magnetization due to the effect 
of electron bound states is smaller than $M_e$ 
by a factor of $\sqrt{\alpha}$. )

We have discussed the magnetization of the axion domain wall.
The origin of the phenomena is in the axion-neucleon ( lepton ) coupling 
which produces 
the attractive potential around the wall. Owing to the potential the fermions
are bounded to the wall with their spins aligned in
the direction perpendicular to the wall. This dynamical mechanism has been 
revealed by the detail calculation. But as we will show, the existence of a 
magnetic field associated with the axion domain wall can be inferred 
from the coupling of the axion and 
electromagnetic field, $g_{\gamma}a\vec{E}\cdot\vec{B}$, where $g_{\gamma}$ is 
a coupling constant. Let us rewrite the coupling in the following,

\begin{equation}
  \int g_{\gamma}\vec{E}\cdot\vec{B}=
\int g_{\gamma}\epsilon_{kji}\partial_jaE_iA_k
\end{equation}
where we have taken the static limit, $\partial B/\partial t =0$; 
$\epsilon_{kji}$ is the antisymmetric tensor and 
$A_k$ is electromagnetic potential. Then we may identify
the electric current $J_k$ associated with the domain wall,

\begin{equation}
J_k=g_{\gamma}\epsilon_{kji}\partial_jaE_i=
g_{\gamma}\epsilon_{3ik}\partial_3aE_i
\end{equation}
This implies a Hall effect; 
the electric current flows
on the wall in the direction perpendicular to 
the electric field. The Hall conductivity\cite{QH}, $\sigma_{xy}$ is given by 

\begin{equation}
\sigma_{xy}=g_{\gamma}\int \partial_3 adx_3=g_{\gamma}f_{PQ}2\pi n
\end{equation}
where $n$ is integer. By noting that 
$g_{\gamma}=\mbox{constant}\times f_{PQ}^{-1}\times e^2/2\pi$ we find that 
the Hall conductivity
is quantized; it reminds us quantum Hall effect\cite{QH}. 
This argument strongly 
suggests the existence of a magnetic field perpendicular to the wall.


Finally as a phenomenological application we 
point out existence of a possible candidate of 
a primordial magnetic field\cite{pri,iwazaki} generated 
by the domain walls. For the purpose we take an axion model\cite{Kim} 
with a heavy charged fermion and with no domain wall problem; the walls
decay quickly. 
The model has axion closed strings 
which surround axion domain walls produced after QCD transition.
Since the heavy fermion has zero modes\cite{Jackiev} on the strings, 
the strings are 
superconducting\cite{Witten}. 
These strings begin to carry a current $J_0=Ml$ 
associated with magnetization $M$,
when the walls are produced; $J_0\sim 1A$ at $\beta=100$ MeV 
with $m_a=10^{-5}eV$. 
This current generates a magnetic field with 
strength $J_0/R_0$ and with length of coherence $R_0$; $R_0$ is the 
radius of the loop which is the order of the size of holizen 
( $R_0=10^6$ cm at temperature $100$MeV ).
The current increases with the string loop shrinking in such 
a way that $JR=J_0R_0$
because the number of the fermions ( $\propto JR$ ) on the loop is conserved. 
Then the magnetic field $B=J/R$ becomes stronger as the radius $R$ of the loop
becomes smaller.
We may estimate explicitely the strength  
of the magnetic field with a critical size of the coherence $10$ cm\cite{dis} 
at the temperature $100$ MeV; magnetic fields 
with smaller sizes of coherence are dissipated 
due to finite conductivity of the universe and do not remain at 
recombination of photons and electrons. 
We find that $B=10^3$ Gauss with the size of coherence, $10$ cm. 
This magnetic field leads to a magnetic field 
with strength $10^{-13}$ Gauss and size of 
coherence $10^9$ cm at the recombination ( $\sim 1$ eV ). 
This is sufficiently strong to be 
a possible candidate of a primordial magnetic field leading to 
galactic magnetic fields in the present universe.

\begin{flushleft}
Acknowledgements

The author would like to thank Prof. M. Kawasaki 
for useful discussions and staff members of theory division in
Institute for Nuclear Study, University of Tokyo for their
hospitality.
\end{flushleft}





\begin{thebibliography}{99}
\bibitem{Wilczeck}S. Weinberg, Phys. Rev. Lett. 40 (1978) 223,\\
F. Wilczeck, Phys. Rev. Lett. 40 (1978) 279.
\bibitem{Pecci}R.D. Peccei and H.R. Quinn, Phys. Rev. Lett. 38 (1977) 1440.
\bibitem{t'Hooft}G. t'Hooft, Phys. Rev. Lett. 37 (1976) 8.
\bibitem{Sikivie}P. Sikivie, Phys. Rev. Lett. 48 (1982) 1156.
\bibitem{Turner}E.W. Kolb and M.S. Turner, The Early Universe, Addison-Wesley,
(1990).
\bibitem{Zeld}Ya.B. Zel'dovich, I.Yu. Kobzarev and L.B. Okun, 
Sov. Phys. JETT 40 (1975) 1.
\bibitem{iwazaki}A. Iwazaki, hep-ph/9608448.
\bibitem{kim}J.E. Kim, Phys. Rep. 150 (1987) 1.
\bibitem{Kim}J.E. Kim, Phys. Rev. Lett. 43 (1979) 103,\\
M.A. Shifman, A.I. Vainshtein and V. I. Zakharov, Nucl. Phys. B166 (1980) 493.
\bibitem{Dine}M. Dine, W. Fischler and M. Srednicki, Phys. Lett. B104 
(1981) 199.
\bibitem{QH}S. Girvin and R. Prange, The Quantum Hall Effect 2nd ed. 
Springer-Verlag, (1990),\\
Z.F. Ezawa and A. Iwazaki Phys. Rev. B47 (1993) 7295
\bibitem{pri}C.J. Hogan, Phys. Rev. Lett 51 (1983) 1488;
M.S. Turner and L.M. Widrow, Phys. Rev. D30 (1988) 2743;
T. Vaschaspati, Phys. Lett. B265 (1991) 258;
B. Ratra, Phys. Rev. D45 (1992) 1913;
B. Cheng and A.V. Olinto, Phys. Rev. D50 (1994) 2421;
M. Gasperini, M. Grovannini and G. Veneziano, Phys. Rev. Lett. 75 (1995) 3796;
F.D. Mazzitelli and F.M. Spedalieri, Phys.Rev. D52 (1995) 6694; 
A. Hosoya and S. Kobayasi, preprint TIT/HEP-298/COSMO-57, 1995.
\bibitem{Jackiev}R. Jackiw and P. Rossi, Nucl. Phys. B190 (1981) 681.
\bibitem{Witten}E. Witten, Nucl. Phys. B249 (1985) 557.
\bibitem{dis}K. Enqvist, A.I. Rez and V.B. Semikov, Nucl Phys. B436 (1995) 49.
\end{thebibliography}
\end{document}